\documentclass{article}
\usepackage{graphicx, amsmath, amsfonts, amssymb, amsthm} % Required for inserting images
\usepackage[left=1in, right=1in, top=1in, bottom=1in]{geometry}
\newtheorem{theorem}{Theorem}
\usepackage[authoryear]{natbib}
\usepackage{float}

\title{Optimal trap cropping investments to maximize agricultural yield}

\author{Matthew H Holden}
\date{School of Mathematics and Physics, The University of Queensland, St Lucia, 4072, Australia\\
Centre for Biodiversity and Conservation Science, The University of Queensland, St Lucia, 4072, Australia}
\begin{document}

\maketitle

\begin{abstract}
Trap cropping is a pest management strategy where a grower plants an attractive ``trap crop'' alongside the primary crop to divert pests away from it. We propose a simple framework for optimizing the proportion of a grower’s field or greenhouse allocated to a main crop and a trap crop to maximize agricultural yield. We implement this framework using a model of pest movement governed by trap crop attractiveness, the potential yield threatened by pests, and functional relationships between yield loss and pest density drawn from the literature. Focusing on a simple case in which pests move freely across the field and are attracted to traps solely by their relative attractiveness, we find that allocating 5--20 percent of the landscape to trap plants is typically required to maximize yield and achieve effective pest control in the absence of pesticides. For highly attractive trap plants, growers can devote less space because they are more effective; less attractive plants are ineffective even in large numbers. Intermediate attractiveness warrants the greatest investment in trap cropping. Our framework offers a transparent and tractable approach for exploring trade-offs in pest management and can be extended to incorporate more complex pest behaviors, crop spatial configurations, and economic considerations.
\end{abstract}

\section*{Recommendations for Resource Managers}
\begin{itemize}
    \item Planting a highly attractive ``trap crop'' that lures pests away from the main crop in an agricultural field can control pests, improve yield, and lower reliance on chemical pesticides.
    \item Because trap plants occupy precious space that could otherwise be used to grow harvestable crops, it is important to optimize their share of the field.
    \item Our results suggest that allocating 5--20\% of the field to trap crops may be required to maximize yield, depending on pest pressure and trap plant attractiveness.
\end{itemize}

\section{Introduction}
Pests remain one of the most significant threats to agricultural productivity worldwide, reducing yields and increasing the need for chemical interventions \citep{savary2019global}. However, widespread pesticide use has led to well-documented issues, including the evolution of pest resistance \citep{sparks2015irac}, harm to non-target species and pollinators \citep{tooker2021newer, basu2024pesticide}, and environmental contamination \citep{de2020occurrence}. These concerns have motivated a growing interest in ecologically based pest management approaches that reduce reliance on chemical controls while maintaining productivity \citep{gurr2003multi, cook2007push, angon2023integrated, gonzalez2019habitat}. Among these, \emph{trap cropping} has emerged as a promising, environmentally friendly tactic for manipulating pest behavior to protect the main crop from pests \citep{sarkar2018application}.

Trap cropping is a strategy where a more attractive plant species or variety is deliberately planted to draw pests away from the main crop, thereby reducing damage and improving overall yield \citep{shelton2006concepts}. The effectiveness of trap cropping has been demonstrated in a variety of agricultural systems, including cotton, brassicas, cucurbits, and legumes \citep{adams2017trap, hokkanen1991trap, cavanagh2009trap,lu2009potential}. However, despite its widespread use, practical guidance on how much land to allocate to the trap crop is lacking. Many studies remain empirical, focusing on evaluating the attractiveness of candidate trap plants for specific crop–pest pairs or small-scale field trials, with limited generalization to broader agroecological conditions \citep{shelton2006concepts, edde2006potential,lee2009attraction}. Moreover, quantitative theory linking system parameters to optimal spatial allocation remains underdeveloped.

Mathematical and simulation models have played a critical role in advancing our understanding of trap cropping and related ecological interventions such as intercropping and companion planting. Early models focused on insect pest dispersal and behavioral responses to attractants or traps embedded within cropping systems \citep{perry1984mathematical}. Since then, a range of modeling approaches have been used to explore how pest movement, crop attractiveness, and spatial layout interact to influence trap crop performance \citep{hannunen2005modelling, holden2012designing, fenoglio2017evaluating}. Simulation-based studies have shown that trap effectiveness depends not only on the strength of attraction but also on spatial configuration and pest mobility. For example, models incorporating biological control agents alongside trap crops reveal strong context dependence, with outcomes shaped by both natural enemy dynamics and landscape structure \citep{banks2024modeling, banks2023compatibility, knudsen2015modeling}. Individual-based models (IBMs) have been used to account for fine-scale behavioral heterogeneity and movement biases in response to trap placement \citep{vinatier2012ibm, simon2020crop}, while recent work has examined how plant dispersion and patch shape influence trapping efficiency \citep{banks1999modelling, ahmed2023simulating}. A broader synthesis of movement modeling in cropping systems underscores the importance of aligning spatial designs with pest dispersal mechanisms and highlights trade-offs between model tractability and ecological realism \citep{garcia2021ecological}. While these studies provide rich insight into pest–crop dynamics and demonstrate the potential of spatially targeted interventions, they do not offer general analytic solutions for the optimal proportion of land to devote to trap cropping. Our work aims to fill this gap by proposing a stylized but tractable framework that captures key trade-offs and yields explicit recommendations for growers.

We start by developing a simple mathematical framework to determine the optimal proportion of a grower’s field that should be allocated to a trap crop to maximize total yield. Our framework is intentionally stylized and analytically tractable. Analytic solutions are especially valuable because they (i) improve transparency by making a direct link between assumptions and outcomes easy to interpret and interrogate, (ii) their solutions serve as baselines for validating more complex future simulation models, and (iii) they allow for general insights to be computed and communicated efficiently. The model is built around two central parameters: the proportion of yield that would be lost to pests in the absence of trap cropping and the relative attractiveness of the trap crop, which determines the likelihood that pests are diverted away from the main crop. Both parameters can be estimated from experimental data, expert elicitation, or synthesis studies on pest behavior and damage functions \citep{zehnder2007arthropod, mitchell2000behavior}.

To illustrate the utility of the framework, we analyze a simplified scenario in which pests move freely across the field and select plants purely based on relative attractiveness. This assumption corresponds to highly mobile pests, such as adult moths that lay their eggs in batches over several days or weeks across multiple plants \citep{jones2019movement}. The assumption of high mobility relative to crop area is especially plausible in smaller greenhouse environments, since pests can rapidly traverse the entire planting area \citep{doehler2023landscape}. Under these conditions, we derive explicit expressions for expected yield and characterize the optimal fraction of land to devote to trap cropping. Using plausible parameter values from the literature, we find that between 5–20\% of the field must typically be allocated to trap crops for effective pest control in the absence of pesticides. Interestingly, we show that the relationship between trap crop attractiveness and optimal area is non-monotonic: highly attractive traps require relatively little area; traps with low attractiveness are ineffective even in large numbers; and intermediate levels of attractiveness require the largest investment. Our framework contributes to the growing body of theory guiding sustainable pest management and can be extended to incorporate more complex dynamics such as pest reproduction, spatial heterogeneity, and economic costs.

\section{Framework}
In this work, we consider a scenario with $N$ pests in a field damaging the main crop harvested by the grower. To reduce this damage, the grower may plant a \textit{trap crop} to divert pests, potentially increasing main crop yield.

A fundamental question in implementing trap cropping is how much land to allocate to trap plants and, consequently, how much to assign to the main crop. Trap plants are typically not harvested and therefore do not directly contribute to yield. In fact, they take up space that would otherwise be used for planting productive main crop plants. However, trap crops can indirectly increase the grower’s yield by reducing pest density on the main crop. This creates a tradeoff: growers are reluctant to allocate space to non-producing plants, but desire lower pest densities.

Assume that the yield of a single main crop plant depends on the pest density on that plant. In other words, a plant’s yield is $y(\rho)$, where $\rho$ is the number of pests on that plant. Let $n$ be the total number of plants that can fit in the field. Let $n_c$ be the number of cash (main crop) plants; then $n - n_c$ is the number of trap plants. Since trap plants are not harvested, and assuming all cash and trap plants are otherwise equal, the total yield across the whole field is given by:

\begin{equation}
Y(n_c) = y\left( \rho(n_c) \right) \cdot n_c \label{Y}
\end{equation}
This equation says that total yield is just the average yield of a cash plant, given the average pest density on cash plants, multiplied by the number of cash plants. As the grower increases the number of cash plants, the multiplication by $n_c$ on the right has an increasing effect, but $y(\rho(n_c))$ is more complicated. If trap plants are effective at reducing pest densities on cash plants, then this term, yield per plant, can decrease when the grower adds cash plants through increases in pest density, $\rho$. However, note that $y$ is not guaranteed to decrease with $n_c$ (even though it decreases in $\rho$) because if trap plants are not effective at trapping or retaining pests, then pests are diluted across an increased number of cash plants. With $N$ pests in the field, the average pest density is

\begin{equation}
\rho(n_c) = \frac{x(n_c) N}{n_c}, \label{rho}
\end{equation}
where $x(n_c)$ is the proportion of insects on the cash crop as a function of the number of cash crops. To maximize total yield, one can differentiate equation \eqref{Y}, which results in general insight into the optimal solution. It says the optimal number of cash plants should satisfy,

% \begin{equation}
% \frac{dY}{dn_c} = \frac{dy}{dn_c} \cdot n_c + y(\rho(n_c))
% \end{equation}

\begin{equation}
y(\rho(n_c)) = -\frac{dy}{dn_c} \cdot n_c, \label{deriv}
\end{equation}
provided some conditions on the functions in equations \eqref{Y} and \eqref{rho} hold (see section \ref{section:generalities}). Equation \eqref{deriv} implies that at the optimal number of cash plants, the yield per plant equals the marginal loss in yield caused by adding another cash plant, summed over all plants. Roughly, if you added a cash plant, the yield of that cash plant should equal the reduced yield summed over all plants due to the increased pest density. If the yield of adding another cash plant is greater than the yield loss it would cause due to pests, it would be optimal to increase the number of cash plants in the landscape. Conversely, if the yield from that plant is not greater than the loss it causes, it is optimal to add more trap plants (i.e., decrease the number of cash plants in the landscape).

While flexible, the framework requires specifying two key relationships: $y(\rho)$, and $x(n_c)$. In the following sections, we demonstrate how empirically supported functions from the literature can be used in this framework to achieve practical recommendations for growers deciding how many trap plants to deploy in their agricultural field or greenhouse.

\section{Yield--Pest relationships}\label{section:yield}

The relationship between yield and pest density has been studied both theoretically and empirically in the literature \citep{dhakal2015assessment}. Many candidate relationships have been proposed. However, they all share some common properties. First, there is some assumed maximum yield per plant in the absence of pests, call this maximum plant yield $k$, in other words, $y(0)=k$. Yield is also assumed to decrease monotonically with pest density, $dy/d\rho<0$ for all $\rho>0$. Common functional forms for $y(\rho)$ include linear declines, exponential and reciprocal curves (which drop sharply at low densities and saturate at high densities), and S-shaped curves, where yield remains high at low pest densities, declines rapidly around a threshold, and then saturates again \citep{dhakal2015assessment}.

For our application, we also consider the maximum proportion of yield lost in the absence of trap plants. That is, we parameterize the yield functions such that $y(N/n)=(1-\beta)k$, where $\beta$ is the proportion of yield lost per plant due to $N$ total pests in a field of $n$ plants in the absence of pest management. For example, if $\beta=1$, $N$ pests will wipe out the entire grower’s yield in the absence of trap crops, whereas if $\beta=0.5$, half of the grower’s yield is lost.

\subsection*{Linear yield loss}

If we assume the yield of a single plant decreases linearly with pest density from a maximum yield of $k$ to $(1-\beta)k$ when pest density is $N/n$, we obtain the yield function,
\begin{equation}
y(\rho)=k\left(1-\frac{\beta n}{N} \rho\right).\label{yieldcurvelin}
\end{equation}

\subsection*{Reciprocal yield}

Assuming yield declines reciprocally with pest density, i.e., $y(\rho) = \alpha / (1 + \gamma \rho)$, and parameterizing such that $y(0) = k$ and $y(N/n) = (1 - \beta)k$, we obtain:
\begin{equation}
y(\rho)=\frac{k(1-\beta)N}{(1-\beta)N+\beta n\rho}.\label{yieldcurverec}
\end{equation}

\subsection*{Exponential yield}

Similarly, exponential yield, parameterized in the same way, can be expressed as 
\begin{equation}
y(\rho)=k(1-\beta)^{n\rho/N}. \label{yieldcurveexp}
\end{equation}

\begin{figure}
\centering
\includegraphics[width=7cm]{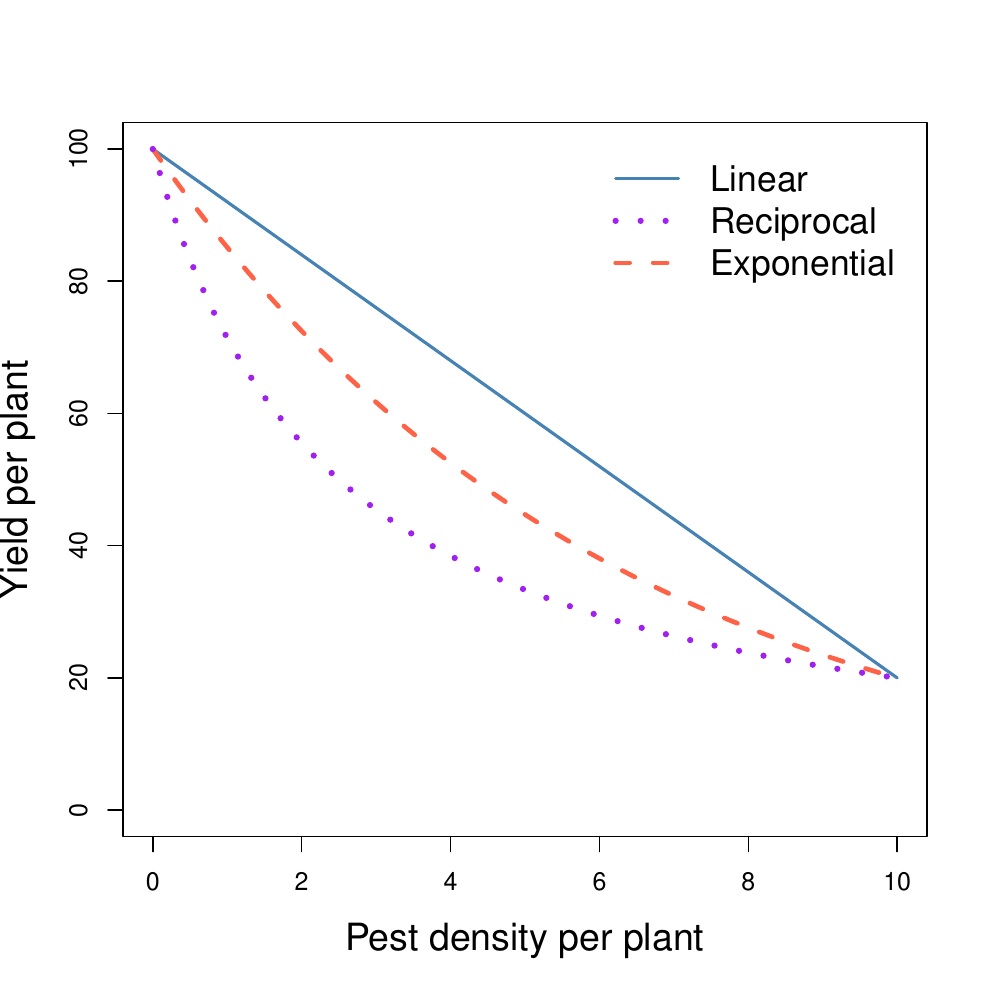}
\caption{Yield curves, $y(\rho)$, as a function of pest density per plant $\rho$, with $N = 1{,}000$, $\beta = 0.8$, $k = 100$, and $n = 10$.}\label{fig:yieldcurves}
\end{figure}

\section{Pest distribution and movement}\label{section:movement}

For simplicity, we assume the pest population is held constant. However, we still require a method to allocate pests between trap and cash plants, since yield depends on both the number of cash plants and the pest load they experience. We describe the pest distribution using a single variable $x_t$, the proportion of the pest population on the main crop (cash plants) at time $t$. Therefore, $1 - x_t$ is the proportion of the pest population on the trap crop (trap plants) at time $t$. Assume the movement dynamics of the pests are described by a simple difference equation.

\begin{equation}
x_{t+1} = f(x_t), \label{difference}
\end{equation}
If $f$ has a stable equilibrium that is reached quickly, this equilibrium can be used in equation \eqref{rho}.

Below, we describe a simple example model first proposed by \citet{holden2012designing} and then use it to compute the optimal allocation of land to cash and trap plants. Consider the case where pests leave plants at each time step with probability $l$. Pests that leave a cash plant settle on another cash plant with probability $s$, while those leaving a trap plant settle on a cash plant with probability $\sigma$. Under these assumptions, the proportion of the pest population on the main crop changes through time iteratively according to the difference equation \eqref{difference} with,

\begin{equation}
f(x) = (1 - l)x + lsx + l\sigma(1 - x). \label{movement}
\end{equation}
The first term accounts for pests that remain on the main crop, the second for those that left and returned, and the third for pests that moved from the trap crop to the main crop.

The equilibrium of this model is

\begin{equation}
x^* = \frac{\sigma}{1 + \sigma - s}, \label{equil}
\end{equation}
which is approximately the proportion of the pest population on the main crop in the long run. Because equations \eqref{difference} and \eqref{movement} define a linear difference equation, the equilibrium is globally asymptotically stable. Therefore, the equilibrium is globally asymptotically stable. There is even an analytic expression for $x_t$, and it can be used to show that solutions, no matter the initial conditions, approach the equilibrium exponentially quickly. Therefore, it suffices to look at equilibrium pest density as a proxy for cumulative pest load as long as crop damage is consistently related to pest density.

So far, our model is simple. We have assumed very little about the spatial structure of the field. The only critical assumption is that all trap plants are treated as identical, and all cash plants are treated identically as well. This assumption implies that pests leaving any trap plant have the same probability of settling on a cash plant, regardless of their starting location. The same applies to pests leaving cash plants. However, this assumption is likely to hold approximately in many systems. If trap plants are planted in rows, for example, from the eyes of a pest on a trap plant, locally, the world looks the same. Note we have yet to introduce any specific model for movement. A specific movement model is needed to determine the settlement probabilities $s$ and $\sigma$.

To specify $s$ and $\sigma$, we assume a common pool dispersal model, in which pests can access all plants each time step. Let $n_T$ be the number of trap plants in the field, and $n_C$ be the number of cash plants. Also, assume a trap plant is $a$ times more attractive than a single cash plant. Since pests view all plants at each time step, their prior location does not affect their settlement decision, and therefore,

\begin{equation}
s = \frac{n_C}{a n_T + n_C} = \sigma. \label{settle}
\end{equation}
In this movement model, the equilibrium in equation \eqref{equil} simplifies to $x^*=s$, providing a simple formula relating the proportion of pests on the cash crop directly to the number of cash crops, $n_C$. Substituting our expression for $x^*$ for $x(n_C)$ then allows us to proceed using the framework to determine the optimal number of cash plants (and hence trap plants) to allocate to the landscape.

It is important to note that in such a simplified, stylized model, we are assuming that pests leave trap plants and cash plants with the same probability, $l$, and trap plants only accumulate more pests than cash plants via their superior attraction once an insect has initiated movement. There is some support for this in the literature for whiteflies as pests of crops in greenhouses \citep{holden2012designing}. 

\section{Optimal number of cash/trap plants to maximize grower yield}\label{section:opt}

Given the movement model described in equations \eqref{difference} -- \eqref{settle} with the simplest yield–pest density relationship, linear yield loss as pest density increases, the optimal number of cash plants in the field to maximize total yield can be derived analytically as

\begin{equation}
n_C^* = \left[\frac{a - \sqrt{\beta a}}{a - 1}\right] n. \label{optnclin}
\end{equation}
To understand this expression, let us first consider the case where $\beta = 1$. Recall that $\beta=1$ means the total number of pests threatening the farm is so great that yield would be zero in the absence of trap crops. In such a case, the optimal number of trap plants, $n-n_C^*$, makes up a high proportion of the landscape. For example, if a trap plant were four times as attractive as a cash plant, the grower would have to devote a third of the landscape to trap plants to maximize yield. Even if the trap crop is 100 times more attractive than the cash crop, the grower would still need to dedicate $9/99$ proportion of the landscape to trap plants (nearly 10 percent). These numbers are quite large as they mean the grower is giving up 10\% of their field to plants that aren’t producing anything they can harvest.

As we decrease $\beta$, this conclusion becomes less severe, but still a substantial proportion of the landscape needs to be devoted to trap plants. For example, even if 40 percent of the grower’s yield is threatened by pests (a typical yield loss due to higher pest densities in organic agriculture), and the trap plant is four times as attractive as the cash plant, then nine percent of the landscape should be devoted to trap plants. If the trap plant is 100 times more attractive, then the optimal allocation of the landscape to trap plants is five percent.

Under nonlinear yield loss functions, such as exponential and reciprocal yield loss, the effects are similar. For reciprocal yield, the optimal number of cash plants is

\begin{equation}
n_C^* = \left[\frac{a(1 - \beta) + \beta - \sqrt{\beta^2 + \beta a(1 - \beta)}}{(1 - \beta)(a - 1)}\right] n, \label{optncrec}
\end{equation}
if the quantity is in $[0,n]$, or $n$ otherwise. For exponential yield, the optimal number of cash plants is

\begin{equation}
n_C^* = \left[\frac{2a - \ln(1 - \beta) - \sqrt{\ln(1 - \beta)(\ln(1 - \beta) - 4a)}}{2(a - 1)}\right] n. \label{optncexp}
\end{equation}
These expressions are more complex than for the linear yield relationship. However, note a key commonality across all three expressions: the optimal number of cash plants in equations \eqref{optnclin} -- \eqref{optncexp} is directly proportional to $n$. This means the bracketed term in each equation can be interpreted as the optimal proportion of the landscape allocated to the cash plant. To visualize how these expressions differ, in the next section, we compare them for all possible parameter combinations.

\section{Illustrative results}\label{section:casestudy}

We start by parameterizing the model using reported quantities available in the literature to form a plausible baseline. For example, it has been shown in choice experiments that 98\% of greenhouse whiteflies, \textit{Trialeurodes vaporariorum}, a common agricultural pest, choose to settle on an eggplant trap plant versus a poinsettia cash plant \citep{lee2009attraction}. This implies that eggplant is 49 times more attractive than poinsettia when the pest has decided to move. It is commonly suggested that without the use of pesticides, a roughly 40 percent reduction in yield is typical in agricultural landscapes \citep{zehnder2007arthropod,oerke2006crop}. We will use these two parameter values as an illustrative baseline, but will also explore the whole parameter space.

\begin{figure}[H]
\centering
\includegraphics[width=15cm]{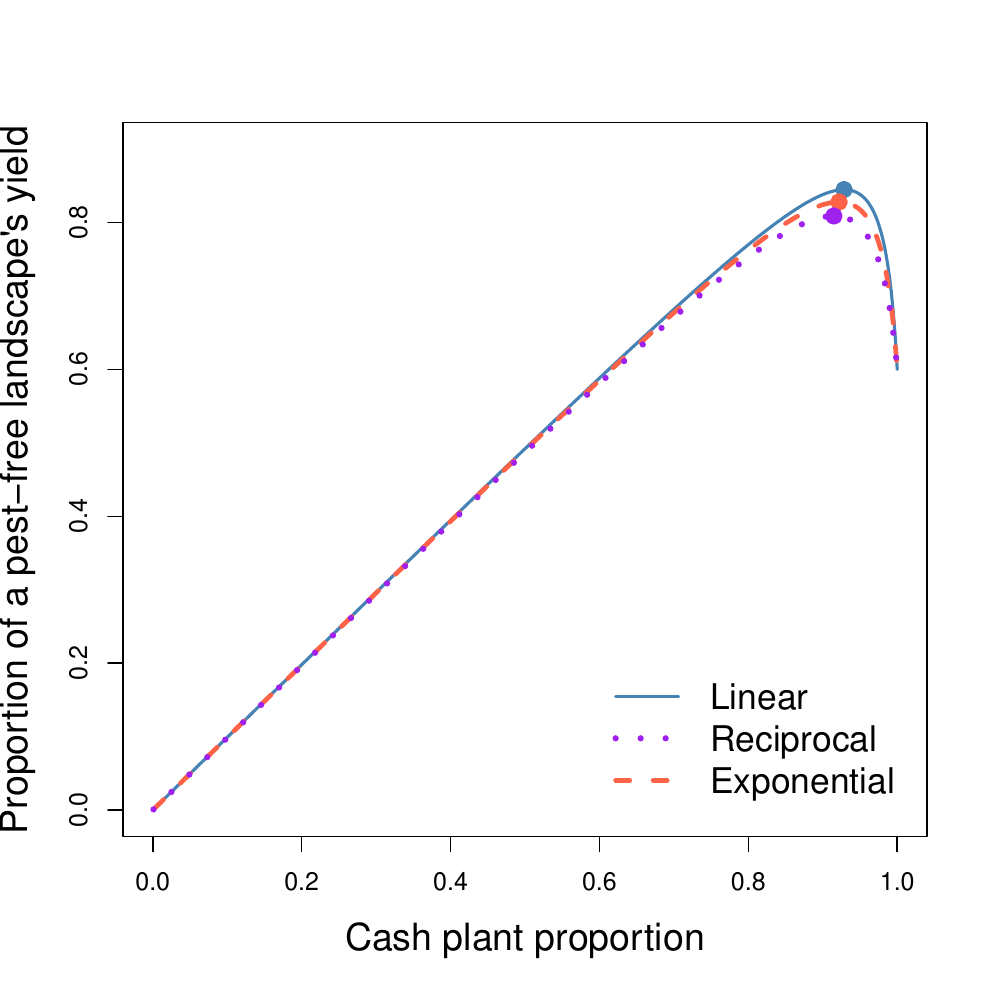}
\caption{Proportion of the total pest-free field's yield (i.e., a field planted only with cash plants) achieved in a system with pests threatening up to 40\% of the yield ($\beta = 0.4$), shown as a function of the proportion of the field devoted to trap plants. Curves correspond to a linear (blue solid), reciprocal (purple dotted), and exponential (red dashed) yield–pest relationship, assuming the trap plant is 49 times more attractive than the cash plant ($a = 49$). Note that when there are no trap plants, the grower loses 40\% of their yield (right endpoint), with steep gains from adding only a few trap plants (moving left). The optimal proportion of the field to devote to trap plants is 7-8.5\%, depending on the yield-pest relationships. Specifically, the optimal cash plant (trap plant) proportion is 92.9\% (7.1\%), 92.2\% (7.8\%), and 91.5\% (8.5\%) for the linear, exponential, and reciprocal yield–pest density relationships, respectively.}\label{fig:yield_nc}
\end{figure}

If $a = 49$ and $\beta = 0.4$, we find that roughly 91.5 -- 92.9 percent of a grower’s greenhouse or field should be allocated to cash plants to maximise yield (Figure \ref{fig:yield_nc}). The nonlinear yield–pest relationships lead to more space being required for trap cropping: 8.5\% for reciprocal yield, equation \eqref{optncrec}, and 7.8\% for exponential yield, equation \eqref{optncexp}, compared to the case when yield declines linearly with pest density (7.1\% trap plants). This occurs because the yield loss per pest is steeper at low pest densities under these relationships, so even small increases in pest density across many cash plants have a large negative impact on total yield.
%%%%%%%%%%%%%%%%%%%%%%%%  FIg 2 - baseline

In the case where there are all trap plants and no cash plants, every additional cash plant adds approximately an additional yield of $y(\rho)$. In other words, in a landscape of no cash plants, adding a cash plant generates a pest-free plant’s worth of yield, because all the pests are on the many trap plants. This explains the linear increasing relationship between total yield and the proportion of the landscape with cash plants for low cash plant proportions (see bottom left of Figure  \ref{fig:yield_nc}). Once the proportion of cash plants begins to exceed 80 percent, then the incremental yield of an additional cash plant starts to be reduced by the associated increased pest density across all cash plants in the system, with an optimal cash plant proportion of 91.5--92.9 percent. At approximately 95 percent of the landscape devoted to cash plants, each additional cash plant severely reduces yield compared to an equivalent investment in trap plants due to pest damage (see the rapid drop off in the right of Figure \ref{fig:yield_nc}).

\begin{figure}[H]
\centering
\includegraphics[width=6.7cm]{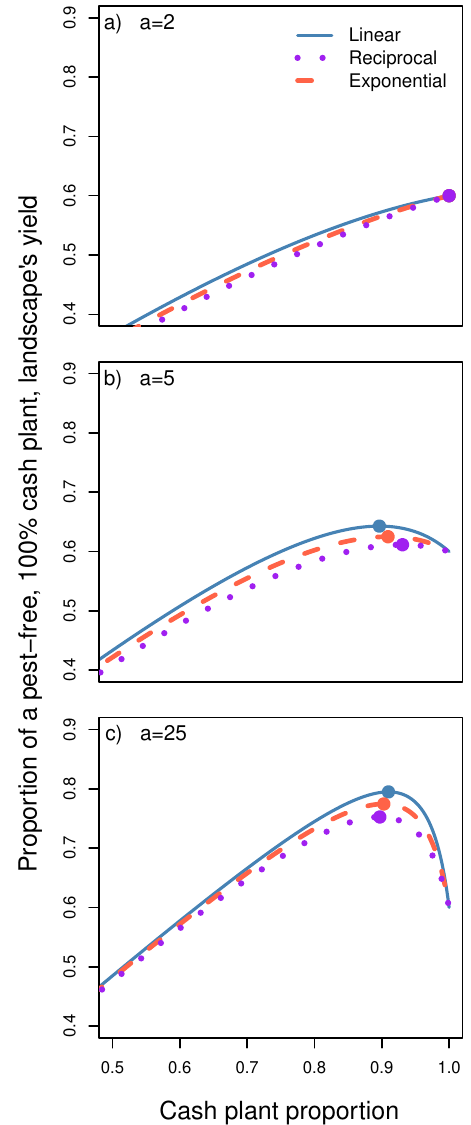}
\caption{Grower yield as a function of the proportion of cash plants for trap cropping systems with less attractive trap plants than the baseline, where trap plants are (a) twice, (b) five times, and (c) 25 times more attractive than the cash plant. For a trap plant that is only twice as attractive as a cash plant, trap cropping is ineffective, and the grower should plant only cash plants, accepting the yield loss caused by pests. For more attractive trap plants, 5 and 25 times more attractive, approximately 7--10\% of the field should be allocated to trap plants.}\label{fig:a}
\end{figure}

However, $a=49$ represents an extremely attractive plant. The question remains: What happens when the trap plant is less attractive? In Figure \ref{fig:a}a, we see that for $a=2$, and $\beta = 0.4$, trap cropping is entirely ineffective at increasing yield. This is because in such cases, the pest is only weakly drawn towards trap plants, and therefore, the reduction in pest pressure on cash plants is minimal; the lost yield from sacrificing land to unharvestable trap crops outweighs the small gains from pest control.
%%%%%%%%%%%%%%%%%%%%%%%%  FIg 3 - attraction

As attractiveness is increased further to five, meaning a pest is five times more likely to settle on a trap plant than a cash plant, it is optimal to devote roughly ten percent of the land to the trap crop (Figure \ref{fig:a}b). However, differences in prevented yield loss are less sensitive to the number of trap plants than for more attractive trap plants (compare Figure \ref{fig:a}b to Figure  \ref{fig:yield_nc} and Figure \ref{fig:a}c).

Interestingly, from figures  \ref{fig:yield_nc} and 3, the optimal number of trap plants is bigger for intermediate attractiveness values of 5 and 25 than for 49. This is because intermediate attractiveness is sufficient to divert pests but not highly efficient, so more area is needed to sufficiently reduce pest pressure and maximize yield. This suggests, counterintuitively, that growers using intermediately attractive trap plants may need to invest more land in trap cropping than those using highly attractive varieties. The figures also show that the nonlinear yield-pest relationships (red and purple) can increase the optimal number of trap crops or decrease it compared to the linear pest-density model, depending on the attractiveness parameter (the order of the colored circles, representing optimal yield, changes when comparing Figure \ref{fig:a}b and to Figure \ref{fig:a}c). These reversals reflect that, at intermediate attractiveness levels, nonlinear damage responses can either amplify or buffer the effects of pest distribution. This depends on which part of the damage curve is most influential.

\begin{figure}[H]
\centering
\includegraphics[width=\textwidth]{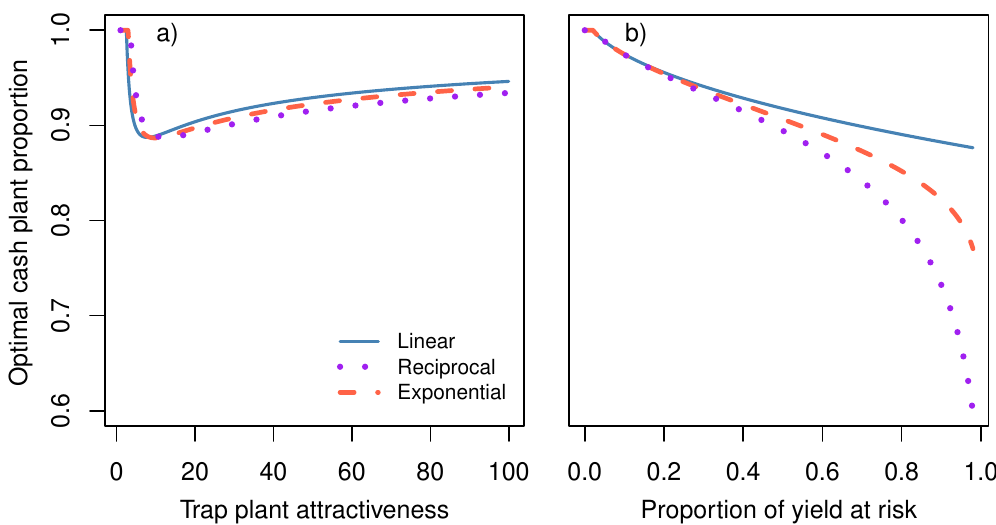}
\caption{The optimal proportion of the landscape to allocate to the main crop (cash plants) as a function of (a) trap plant attractiveness and (b) the proportion of yield at risk from pests. For intermediately attractive trap plants (5--20 times as attractive as the cash plant), more than 10\% of the landscape should be sacrificed to trap plants, as they provide sufficient efficacy to achieve pest reduction, but are not effective enough to work at low densities. For low attraction ($a < 4$), trap cropping is ineffective. For highly attractive trap plants, their efficacy is high enough that a smaller proportion of the landscape can achieve the necessary reduction in pest densities to maximize yield.}\label{fig:optab}
\end{figure}

To investigate this further, we plotted the optimal proportion of cash plants in equations \eqref{optnclin} -- \eqref{optncexp} versus attractiveness, for the baseline yield at risk of 40 percent (Figure \ref{fig:optab}a). When trap plants are between one and 2.6 times more attractive than the main crop, the whole field should be cash plants, as trap cropping is not effective enough to make up for lost yield. From an attractiveness of 3.6 to eight, the optimal proportion of cash plants declines rapidly to a minimum of $88$ percent of the landscape under all yield-pest density relationships. However, the decline is slightly more rapid for the linear case, where the 88 percent minimum occurs at an attractiveness of eight, compared to the exponential and reciprocal cases, where the minimum is achieved at higher attractiveness values of 10 and 12.5, respectively. After achieving the minimum, for larger attractiveness values, the optimal proportion of the landscape devoted to cash plants gradually rises again. This is because, as trap crops become highly effective, less land allocation is required to achieve adequate pest control. For even nearly perfect attractiveness, e.g., 100 times more attractive than the cash crop, the grower still needs to devote five, six, and seven percent of the landscape to the trap crop, for the linear, exponential, and reciprocal relationships, respectively.   

%%%%%%%%%%%%%%%%%%%%%% Fig 4: optimal nc vs. a and b

Optimal trap cropping is even more sensitive to the maximum yield at risk due to pests, $\beta$, than the attraction parameter. We plotted the optimal proportion of cash plants in equations \eqref{optnclin} - equation \eqref{optncexp} versus yield at risk, $\beta$, for the baseline trap plant attractiveness of 49 percent (Figure \ref{fig:optab}b). In general, the optimal proportion of the landscape dedicated to the cash crop declines monotonically as there is more yield at risk due to high pest densities (see decreasing trend in Figure \ref{fig:optab}b). This is particularly more severe for the more nonlinear pest yield relationships. 

\begin{figure}[H]
\centering
\includegraphics[width=6.5cm]{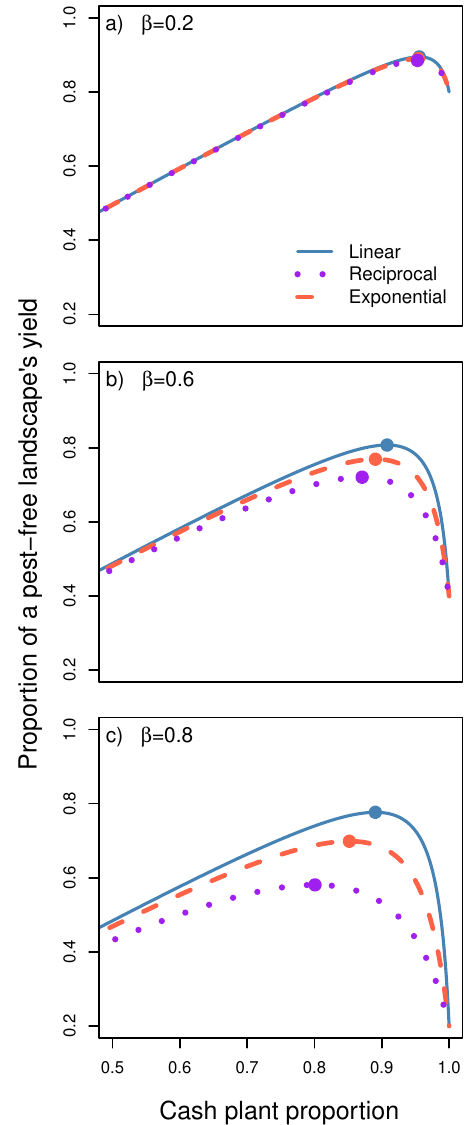}
\caption{Grower yield as a function of the proportion of cash plants for trap cropping systems with different levels of yield at risk compared to the baseline in Figure  \ref{fig:yield_nc}: when (a) 10\%, (b) 30\%, and (c) 60\% of maximum yield is at risk due to damage from pests.}\label{fig:b}
\end{figure}

For the reciprocal yield relationship, this is particularly severe. If 100\% of the yield is at risk, the cash crop should make up roughly 60\% of the landscape to maximize yield. For high, but more typical, yield at risk, such as 60 and 80 percent, optimal cash crop allocations correspond to $87$ and $80$ percent, respectively. For linear and exponential relationships, the trend is similar but less severe. The relationship is strongest for the reciprocal yield curve because even low pest density causes a steep drop-off in yield. Therefore, more trap plants are required to achieve optimal yield. The exponential is an intermediate case between this severely nonlinear yield–pest relationship and the more gradual, linear, pest–yield relationship. However, even in the linear case, if 100 percent of the yield is at risk, still 88 percent of the field needs to be devoted to cash plants to maximize yield. To see the differences between the yield achieved at all cash crop proportions for different yields at risk, see Figure \ref{fig:b}.

%%%%%%%%%%%%%%%%%%%%%% Fig 5: beta

\begin{figure}[H]
\centering
\includegraphics[width=\textwidth]{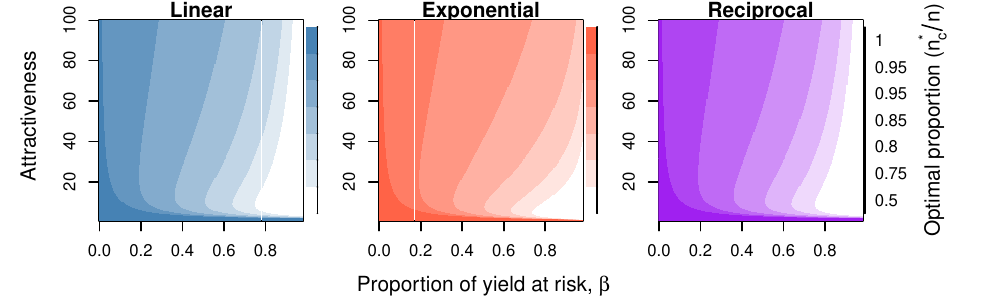}
\caption{Heatmap of the optimal proportion of the landscape to allocate to a cash crop across a grid of yields-at-risk, $\beta$, and trap attractiveness, $a$. Darker regions represent greater investment in cash crops (low trap crop area). The white and lightest regions indicate 25–50\% of the landscape is allocated to cash crops, implying substantial trap cropping. In most of the parameter space, optimal trap cropping investments require 5–20\% of land area, as the white and next lightest regions, along with the darkest region, are small.}\label{fig:heatmap}
\end{figure}

In the previous plots, we have fixed one parameter to be at the baseline value while varying the other. A complete sensitivity analysis of the optimal cash crop allocation across all possible combinations of attractiveness and yields at risk is displayed as a heatmap in Figure \ref{fig:heatmap}. Across all yield-pest density relationships, yields at risk, and attractiveness, the vast majority of the parameter space corresponds to the case when 80- 95 percent of the landscape should be dedicated to the cash crop to maximize agricultural yield. Low optimal cash plant proportions of less than five percent only occur when less than 30 percent of the grower's yield is at risk (left second darkest region). Trap cropping should be avoided only when trap plants are both unattractive and low yields are at risk (bottom-left dark regions in all panels)

%%%%%%%%%%%%%%%%%%%%%% Fig 6: multi plot

The figure also demonstrates the result that intermediate attraction warrants the greatest trap cropping investment, compared to highly attractive and unattractive trap crops, is general across parameterizations. That is for all yield-pest dentistry relationships (all panels) and parameter values for yields at risk (horizontal axis), as long as the trap crop is attractive enough to warrant some investment, the cash plant allocation that maximizes yield is highest for trap plants that are only a few times more attractive than the cash plant. To see this, examine vertical transects through each figure. Note there is a dark region on each side of the transect and a lighter region for intermediate attraction values for a wide range of yields at risk. 

\section{Generalities: existence of a single optimal cash plant proportion}\label{section:generalities}

In sections \ref{section:yield} -- \ref{section:casestudy}, we focused on a specific model of pest movement and crop yield to demonstrate the framework and improve intuition. This required specifying several functional relationships. The question remains whether such results are possible across a wide range of functional forms and models. Below, we present general conditions on these functions, such that they guarantee a unique optimal number of cash plants (a sufficient condition).

\begin{theorem}\label{thm}
There is a unique optimal number of cash plants, $n_C^* \in (0,n]$, satisfying equation \eqref{deriv}, that maximizes total yield, $Y(n_C)$, if the following three properties hold:
\begin{enumerate}
    \item The function $x$ is positive, continuously differentiable, and strictly increasing on $(0,n)$.
    \item The function $E$, defined by
    \begin{equation}
        E(n_C) = \frac{n_C}{x(n_C)} \frac{dx}{dn_C} - 1, \label{elasticity}
    \end{equation}
    is positive and strictly increasing with respect to $n_C$ on $n_C\in(0,n)$.
    \item The function $y$ is continuously differentiable, and  $y'/y$ is monotonic in $\rho$.\\
\end{enumerate}
\end{theorem}
\noindent Below, we provide a biological explanation of the theorem. However, see the appendix for a rigorous proof. The intuition behind condition 1 is simple: As you increase the number of cash plants (and therefore decrease the number of trap plants), it requires that the density of pests per plant increases. In other words, trap plants have to reduce pest density on the cash plants as you add more trap plants to the landscape. This is almost surely satisfied for any trap cropping system; otherwise, the trap plant is ineffective and would not be considered by the grower.

Note that equation \eqref{elasticity} in condition 2 is just the elasticity of pest density on a single cash plant with respect to the number of cash plants. This is because
\begin{equation}\label{elastic_derivation}
\frac{n_C}{\rho} \frac{d\rho}{dn_C} = \frac{n_C}{x(n_C)} \frac{dx}{dn_C} - 1 := E(n_C).
\end{equation}
Elasticity, in economics, is the proportional rate of change of one quantity with respect to a proportional change in the other quantity. For example, if the elasticity in equation \eqref{elastic_derivation} is two, then, approximately, a one percent increase in cash plants will cause a two percent increase in the proportion of pests on a cash plant. Therefore, condition 2 says that the elasticity of pest density on a single plant with respect to the number of cash plants present should be an increasing function.

To provide intuition for condition 2, remember that total yield across the field is the product of yield per cash plant times the number of cash plants, as seen in equation \eqref{Y}. Consider the case where there are only 100 cash plants. In this case, a one percent boost in cash plants only adds a single plant, increasing yield by the yield from only that one plant, roughly $y(\rho)$. Now, in the case where there are 1,000 cash plants, increasing the number of plants by one percent increases yield by roughly ten times $y(\rho)$. Because yield scales with the number of cash plants, any yield loss per plant resulting from additional cash plants must be steep enough to offset the benefits of increasing plant numbers

Since we have already assumed that the yield of a plant decreases as you increase the density of the pests (see section \ref{section:yield}), condition 3 of the theorem just restricts the behavior of this decrease. It requires that either every additional individual pest added to a plant decreases yield more than the one before, or each pest decreases yield less than the one before, but it never switches back and forth depending on how many pests are on the plant.

Theorem \ref{thm} means that for many mathematical models of pest movement and yield-pest relationships, one can set the derivative of $Y$ to zero and solve for the unique optimal number of cash and trap plants, either analytically or numerically, without needing any global optimization algorithms. It is easy to show that the model of pest movement in equations \eqref{difference} -- \eqref{equil} and all proposed yield-pest density relationships in equations \eqref{yieldcurvelin} -- \eqref{yieldcurveexp} satisfy the conditions in Theorem \ref{thm}. However, a yield-pest density curve with an inflection point will fail to satisfy condition 3, and therefore, multiple local optima will need to be checked in such a model. 

\section{Discussion}
Our results show that trap cropping can substantially improve agricultural yield, but only under specific conditions. Using a simple yet flexible framework, we derived analytic expressions for the optimal number of cash and trap plants, demonstrating how optimal strategies depend critically on two parameters: the proportion of yield at risk due to pests ($\beta$), and the relative attractiveness of the trap crop compared to the cash crop ($a$). When trap crops are only marginally more attractive than the cash crop (e.g., $a = 2$), trap cropping fails to improve yield, even under extreme pest pressure. In contrast, when trap crops are highly attractive (e.g., $a = 49$), yield gains can be substantial even with relatively small land allocations to trap plants.

Our theoretical results provide general conditions under which a unique optimal solution for the number of cash plants exists. In particular, the condition that the elasticity of pest density on a cash plant with respect to the number of cash plants is increasing (Condition 2 in Theorem~1) plays a central role. This elasticity captures how pest distribution responds to changes in landscape composition and connects directly to the efficiency of trap cropping. These results complement other theoretical work on pest suppression strategies that use elasticity-based reasoning, in diverse fields from agricultural economics to invasive species management \citep{headley1972economics, sanchirico2010spatial,holden2016economic}.

Our findings have clear practical relevance. First, they demonstrate that highly attractive trap crops can offer strong control with minimal land sacrifice. However, even the most attractive trap crops require 5–10\% of the landscape to be devoted to non-harvestable plants when pest pressure is severe. This finding is particularly important for organic and low-input systems, where pesticides are avoided and trap cropping is one of few viable pest management options \citep{shelton2006concepts}. Second, the analysis emphasizes that blanket recommendations about trap crop allocation are unlikely to be effective. Instead, effective implementation requires system-specific knowledge: the value of $\beta$ and $a$ must be estimated for each crop–pest combination. Our formulas provide a valuable tool for translating this information into actionable trap cropping recommendations.

As with any modeling approach, our framework makes simplifying assumptions. We assume a fixed pest population ($N$ constant), homogeneous mixing of pests across the landscape, and identical cash and trap plants aside from attractiveness. These assumptions ignore important ecological complexities, such as spatial heterogeneity in plant placement within a field, pest dispersal behaviors, and pest reproduction \citep{vasquez2017trap}. Furthermore, we consider an agricultural field or greenhouse in isolation, ignoring its place within a landscape of natural areas, roads, and other agricultural fields, where optimal strategies may depend on neighboring pest management strategies \citep{drechsler2024game,lampert2025eradicating}. Additionally, our model also treats trap plant effectiveness solely in terms of attractiveness, without considering pest arrestment, retention, or mortality on trap plants. In systems where trap plants kill or immobilize pests (e.g., through glandular trichomes or nectar-feeding traps), \citep{cook2007push} trap cropping will likely be more effective, and our model's recommendations may not hold.  

Nonetheless, our framework is both general and flexible.  We chose to demonstrate it with a deliberately simple and analytically tractable model, designed to clarify the fundamental trade-offs in trap cropping and to serve as a theoretical foundation for future extensions. By distilling the problem to its essential elements, the model provides general insights that remain easily interpretable and transferable. Specifically, the fact that analytic solutions are achieved and only depend on two parameters allowed us to explore the whole parameter space, presenting a complete analysis. This means future researchers adapting the model to incorporate more complex mechanisms have a concrete baseline that they can robustly compare the results to.

Several avenues merit further exploration. Spatially explicit models could help examine how the arrangement of trap and cash plants influences pest dynamics, building on work in habitat fragmentation and landscape ecology \citep{margosian2009connectivity,martin2016scale}. Stochastic models could incorporate variable pest dispersal patterns or plant preferences.  Further, while our framework was designed for trap cropping, other systems where the grower sacrifices space via companion planting to increase yield in the rest of the field may also work for our framework. For example, a modified version of the model could be used to optimize the amount of area devoted to companion plants that attract beneficial organisms such as pollinators, which, unlike pests, enhance rather than reduce crop yield \citep{albrecht2020effectiveness,stein2017bee, blaauw2014flower}. In this context, the spatial trade-off between yield-producing and service-providing plants remains, but the ecological mechanism is reversed. Such applications may help inform broader ecological management strategies in agriculture.

Trap cropping remains a promising pest management strategy, especially for growers seeking to minimize chemical inputs. However, its success depends critically on the amount of space the grower allocates to trap plants. Our framework offers a tractable approach for identifying optimal land allocations. As such, it provides a foundation for both future theoretical exploration and practical decision-making in sustainable pest management.

\bibliographystyle{apalike} 
\bibliography{traprefs}
\section{Appendix}
The following is a proof of Theorem \ref{thm} in section \ref{section:generalities}, which provides conditions for when there is a unique optimal cash crop allocation as a proportion of the landscape.
\begin{proof}
Assume conditions 1-3 hold. From equations \eqref{Y} and \eqref{rho}, and condition 1, $Y$ is continuously differentiable on $(0,n]$. We, therefore, need only show that $dY/dn_C$ has at most one sign change on the interval $[0,n_C]$. Note that,
\begin{align*}
\frac{dY}{dn_c} &= \frac{dy}{d\rho}\frac{d\rho}{dn_C}  \cdot n_c + y(\rho(n_c))\\
&= y(\rho(n_c)) \left[\frac{dy}{d\rho}\frac{d\rho}{dn_C}  \cdot \frac{n_c}{y(\rho(n_c))} + 1\right]\\
&= y(\rho(n_c)) \left[1-E(n_C)F(\rho)\right],
\end{align*}
where $F(\rho)=-\rho y'/y$. By condition 3, $y'/y$ is monotonic in $\rho$. Further, $\rho$ is increasing in $n_c$ by condition (2). To see why, recall that condition 2 and equation \eqref{elastic_derivation} imply,
\begin{equation*}
\frac{n_C}{\rho} \frac{d\rho}{dn_C} > 0.
\end{equation*}
Since $\rho$ and $n_c$ are positive, $d\rho/dn_C > 0$. Thus, $y'/y$ is the composition of a monotonic and strictly increasing function, and therefore monotonic in $n_C$. Thus, $F$ is monotonic, as it is proportional to the product of a monotonic and strictly increasing function. Since $y(\rho(n_c))$ is positive, and $E(n_C)F(\rho)$ is monotonic, $dY/dn_C$ has at most one sign change on $(0,n)$.
\end{proof}
\vspace{1em}

\end{document}